\documentclass{llncs}

\usepackage{textcomp} 
\usepackage{amstext} 
\usepackage{graphicx}
\usepackage[pdftex,
      colorlinks=true,
      urlcolor=blue,       
      filecolor=green,     
      linkcolor=red,       
      pdftitle={Papers by AUTHOR},
      pdfauthor={Your Name},
      pdfsubject={Just a test},
      pdfkeywords={test testing testable},
      pagebackref,
      pdfpagemode=None,
      bookmarksopen=true]{hyperref}

\newcommand{\muon}		{\mbox{$\mu $ }}

\begin{document}

\title{Design, status and perspective of the Mu2e crystal calorimeter}

\author{G.~Pezzullo\inst{2}, N. Atanov\inst{3}, V. Baranov\inst{3}, J. Budagov\inst{3},
  F. Cervelli\inst{2}, F. Colao\inst{5}  , E. Diociaiuti\inst{5},
  M. Cordelli\inst{5}, G. Corradi\inst{5}, E. Dan\`e\inst{5},
  Yu. Davydov\inst{3},
  S. Donati\inst{1,2}, R. Donghia\inst{5}, S. Di Falco\inst{2},
  B. Echenard\inst{4}, L. Morescalchi\inst{2},
  S. Giovannella\inst{5}, V. Glagolev\inst{3},
  F. Grancagnolo\inst{6},
  F. Happacher\inst{5}, D. Hitlin\inst{4}, M. Martini\inst{5, 8},
  S. Miscetti\inst{5} , T. Miyashita\inst{4}, L. Morescalchi\inst{2},
  P. Murat\inst{7}, 
  E. Pedreschi\inst{2}, F. Porter\inst{4}, F. Raffaelli\inst{2}, M. Ricci\inst{5,8},
  A. Saputi\inst{5},
  I. Sarra\inst{5, 8},
  F. Spinella\inst{2}, G. Tassielli\inst{6}, V. Tereshchenko\inst{3},
  R. Y. Zhu\inst{4}}
  
\institute{
  Departement of Physics, University of Pisa, Largo B. Pontecorvo 3, Pisa , Italy \and
  INFN sezione di Pisa, Italy, Largo B. Pontecorvo 3, Pisa , Italy \and
  Joint Institute for Nuclear Research, Joliot-Curie 6, Dubna, Russia \and
  Departement of Physics, California Institute of Technology, 1200 E California Blvd, Pasadena (CA), USA \and
  INFN Laboratori Nazionali di Frascati, via Enrico Femri 40, Frascati, Italy \and
  INFN sezione di Lecce, Via Arnesano 73100, Lecce, Italy \and 
  Fermi National Accelerator Laboratory, Main Entrance Rd, Batavia (IL), USA \and
  Department of Energy, University Guglielmo Marconi, via Plinio, 44, 00193 Roma, Italy
}

\date{}
\maketitle
\email{pezzullo@pi.infn.it}

\begin{abstract}
 The Mu2e experiment at Fermilab will search for the charged lepton
 flavor violating process of neutrino-less $\mu \to e$ coherent
 conversion in the field of an aluminum nucleus. Mu2e will reach a
 single event sensitivity of about $2.5\cdot 10^{-17}$ that
 corresponds to four orders of magnitude improvements with respect to
 the current best limit. The detector system consists of a straw tube
 tracker and a crystal calorimeter made of undoped CsI coupled with
 Silicon Photomultipliers. The calorimeter was designed to be operable
 in a harsh environment where about 10 krad/year will be delivered in
 the hottest region and work in presence of 1 T magnetic field. The
 calorimeter role is to perform $\mu$/e separation to suppress cosmic
 muons mimiking the signal, while providing a high level trigger and a
 seeding the track search in the tracker. In this paper we present the
 calorimeter design and the latest R$\&$D results.

\end{abstract}

\section{Intorduction}
Observation of the neutrino oscillation during the last decades
boosted the interest of the experimental comunity in the search of
Lepton Flavour Violating (CLFV) processes also in the field of charged
leptons. 
The muon conversion represents a powerful
channel to search for CLFV, because it is characterized by a
distinctive signal consisting in a mono-energetic electron with energy
$\rm E_{\mbox{ce}} = \mbox{m}_{\mu} - \mbox{E}_b - \mbox{E}_\mu^2/(2\mbox{m}_{\rm N})$ ,
where $\mbox{m}_{\mu}$ is the muon mass at rest, $\mbox{E}_b \sim
\mbox{Z}^2\alpha^2\mbox{m}_{\mu}/2$ is the muonic atom binding energy
for a nucleus with atomic number $\mbox{Z}$, $\mbox{E}_\mu$ is the
nuclear recoil energy, $\mbox{E}_\mu = m_\mu - \mbox{E}_b$, and
$\mbox{m}_{\rm N}$ is the atomic
mass~\cite{annurev.nucl.58.110707.171126}. In case of aluminum, which is the
major candidate for upcoming experiments, E$_{ce}= 104.973$
MeV~\cite{PhysRevD.66.096002}. 



\section{Calorimeter design}
The calorimeter consists of two disks anulii, separated by 75 cm, with
ineer (outer) radius of 37.4 (66) cm. Each disk is filled wih 678
undoped CsI crystals 20 x 3.4 x 3.4 cm$^3$. The inner region is left
un-instrumented to avoid interactions with low energy electrons, while
the separation between the disks maximize the acceptance for the
conversion electrons (CE). Each crystal is read out by two arrays 2x3 of
6x6 mm$^2$ SiPM.  Each SiPM array is matched to a front end
electronics board (FEE) that provides an amplification stage and also
local voltage regulation. Signal form the FEE is then digitized by a
custom made waveform digitizer @ 200 Msps~\cite{Mu2eWD}.
\begin{figure}[h!]
  \centering
  \includegraphics[width=0.4\textwidth] {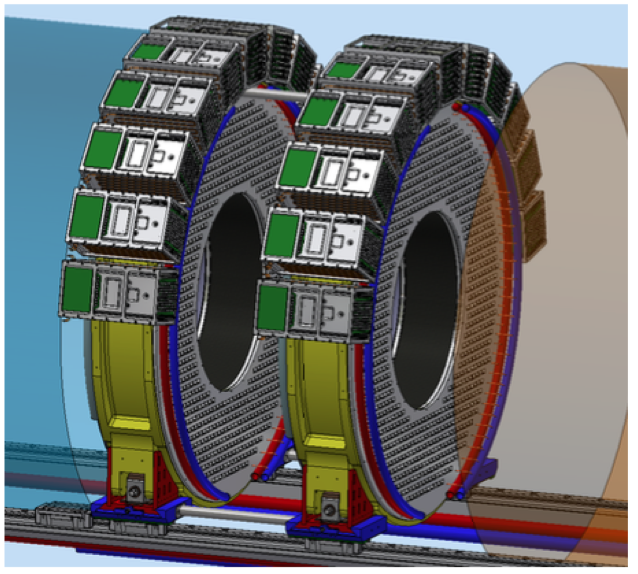}
  \includegraphics[width=0.22\textwidth] {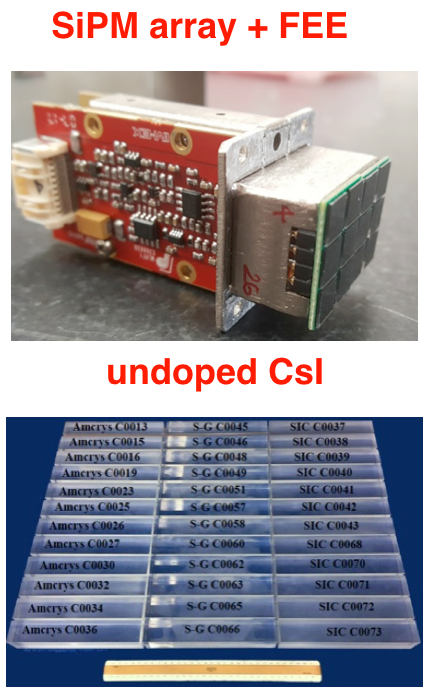}
  \caption{Left: Calorimeter design. Right: Undoped CsI
    crystals from the pre-production and one SiPM + FEE module.}
  \label{fig:design}
\end{figure}
Figure~\ref{fig:design} shows a 3D representation of the calorimeter
and a picture of the crystals and SiPM from the pre-production. A more
detailed description of the calorimeter desgin can be found on
reference~\cite{CaloRef}.

\section{Role of the caloriemter in Mu2e}
\begin{figure}[h!]
  \centering
  \includegraphics[width=0.8\textwidth] {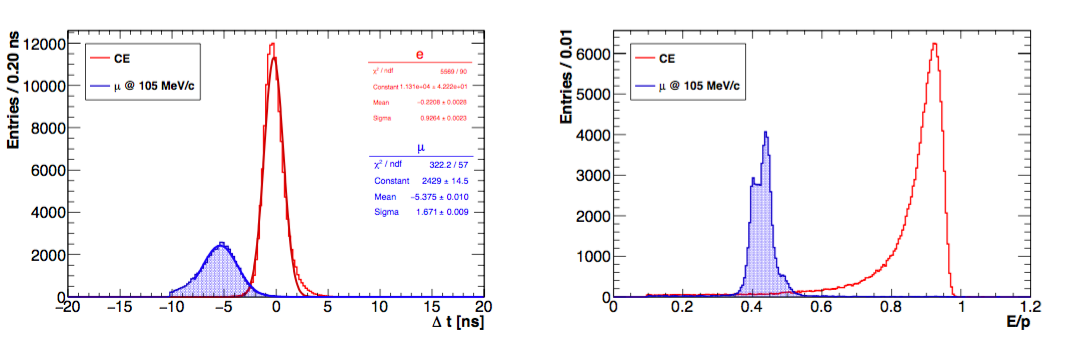}
  \caption{Left: distribution in $\Delta t$ for CE and \muon @ p=105
    MeV/c. Right: Distribution in E/p for CE and \muon @ p=105 MeV/c.}
  \label{fig:PID}
\end{figure}

\begin{figure}[h!]
  \centering
  \includegraphics[width=0.8\textwidth] {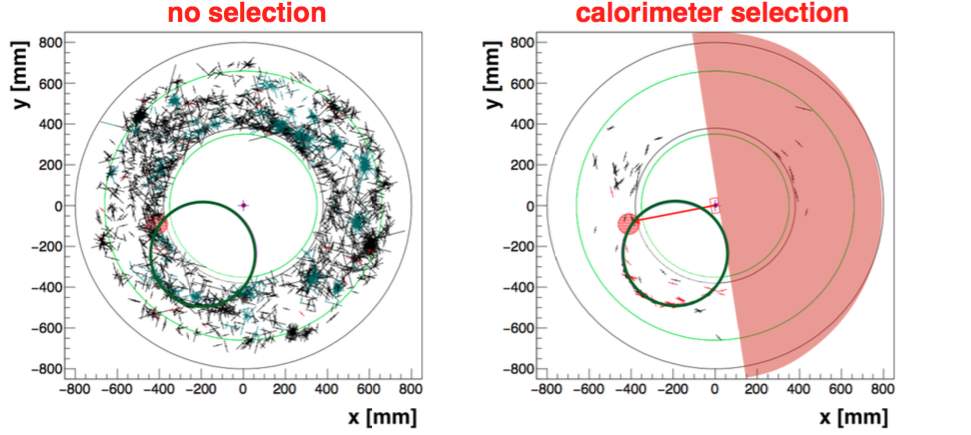}
  \caption{Transverse view of an event display for a CE event with
    background hits included, with (right) and without (left) the
    calorimeter pre-selection. The black crosses represent the straw
    hits, the red bullets the calorimeter clusters, and the red circle
    the CE trajectory.}
  \label{fig:CALOSEED}
\end{figure}

The main role of the calorimeter in Mu2e is to provide particle
identification capabilities that are essentials to distinguish Cosmic
\muon @ p=105 MeV/c mimicking the CE. The rejection algorithm is
based on a likelihood ratio that uses as input the following
observables: 1) E/p: ratio of the track momentum and the calorimeter
cluster energy; 2) $\Delta t$: time residual between the calorimeter
cluster and the track time as extrapolated to the
calorimeter. Figure~\ref{fig:PID} shows the distribution of E/p and
$\Delta t$ for CE and \muon @ p=105 MeV/c.
The calorimeter information are also useful for driving the track
search in the tracker by means of time and spatial correlations with
the hits in the chamber associated to the same particle that produced
the calorimeter cluster~\cite{CaloSeed}. Figure~\ref{fig:CALOSEED}
shows how the hits pre-selection reduces the number of background hits
in a typical event with one CE overlaid with the expected background.
\section{R \& D}
\begin{figure}[h!]
  \centering
  \includegraphics[width=0.55\textwidth] {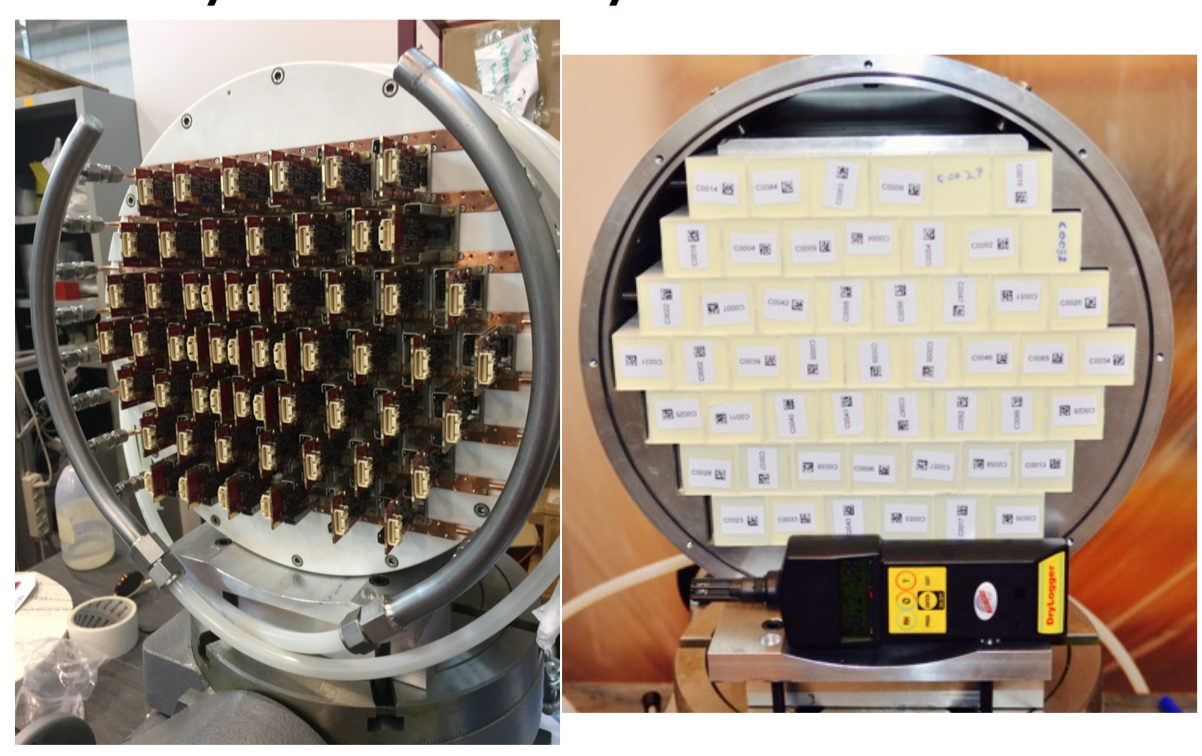}
  \caption{Calorimeter prototype during the assembly phase.}
  \label{fig:module0}
\end{figure}

During late 2016 we started the pre-production of the crystals, SiPM
and FEE boards that were used for the assembly of the final
calorimeter prototype, see figure~\ref{fig:module0}. This prototype
has a key role in validating the expected physics performance and also
check several mechanical properties, like the performance of the
cooling system and the assembly procedures.
During May 2017 a test beam was performed at the Beam Test Facility in
Frascati (Italy)~\cite{Mazzitelli2003524} using an $e^-$ beam in the
range [60, 120] MeV.

\section{Summary}
In this paper the calorimeter project for the Mu2e experiment has been
presented. We showed that the Mu2e calorimeter design is able to
provide particle identification capabilities, an high level trigger
and is also a helpfull tool for improving the tracking pattern
recognition. A test beam with a prototype of the calorimeter was
performed during May 2017 using an electron beam in the energy range
[60, 120] MeV.

\section*{Acknowledgment}
This work was supported by the US Department of Energy; the Italian
Istituto Nazionale di Fisica Nucleare; the US National Science
Foundation; the Ministry of Education and Science of the Russian
Federation; the Thousand Talents Plan of China; the Helmholtz
Association of Germany; and the EU Horizon 2020 Research and
Innovation Program under the Marie Sklodowska-Curie Grant Agreement
No.690385.


\begin{thebibliography}{[MT1]}
  
  
\bibitem{annurev.nucl.58.110707.171126} Marciano, W. J.  and others,
  Charged Lepton Flavor Violation Experiments, Annual Review of
  Nuclear and Particle Science, 58, 1, 315-341, 2008
  
  
\bibitem{PhysRevD.66.096002} Kitano, R. {\it et al},
  Detailed calculation of lepton flavor violating muon electron
  conversion rate for various nuclei, Phys. Rev., D66, 2002
  


\bibitem{Mu2eWD} S. Di Falco {\it et al} Components Qualification for
  a Possible use in the Mu2e Calorimeter Waveform Digitizer, JINST,
  12, 03, C03088, 2017

\bibitem{CaloRef} Mu2e Calorimeter group, The Mu2e Calorimeter Final Technical Design Report,
  http://mu2e-docdb.fnal.gov/cgi-bin/ShowDocument?docid=8429

\bibitem{CaloSeed} G. Pezzullo and P. Murat, The calorimeter-seeded
  track reconstruction for the Mu2e experiment at Fermilab, 2015 IEEE
  Nuclear Science Symposium and Medical Imaging Conference (NSS/MIC), 1-3, 2015

\bibitem{Mazzitelli2003524} G. Mazzitelli {\it et al}, Commissioning
  of the DA$\phi$NE beam test facility, NIM A, 515, 3, 524-542, 2003
\end{thebibliography}
\end{document}